\documentclass[runningheads]{llncs}
\usepackage{graphicx}
\usepackage{appendix}
\usepackage{hyperref}
\hypersetup{
    colorlinks=true,
    linkcolor=blue,
    filecolor=magenta,      
    urlcolor=blue,
}
\usepackage[perpage]{footmisc}
\begin{document}
\title{Labelling imaging datasets on the basis of neuroradiology reports: a validation study}
\titlerunning{NLP for neuroradiology image labelling}

\author{David A. Wood\inst{1} \and
Sina Kafiabadi\inst{2} \and
Aisha Al Busaidi\inst{2} \and
Emily Guilhem\inst{2} \and
Jeremy Lynch\inst{2} \and
Matthew Townend\inst{3} \and
Antanas Montvila\inst{2} \and
Juveria Siddiqui\inst{2} \and
Naveen Gadapa\inst{2} \and
Matthew Benger\inst{2} \and
Gareth Barker\inst{4} \and
Sebastian Ourselin\inst{1} \and
James H. Cole\inst{4,5} \and
Thomas C. Booth\inst{1,2}\\
\vspace{1em}
MR Imaging abnormality Deep learning Identification study (MIDI) Consortium}

%
\authorrunning{D. Wood et al.}
%
\institute{School of Biomedical Engineering, King’s College London \and
King’s College Hospital, London, UK \and
Wrightington, Wigan \& Leigh NHSFT \and
Institute of Psychiatry, Psychology \& Neuroscience, King’s College London \and
Centre for Medical Image Computing, Dementia Research, University College London}
\maketitle              

\begin{abstract}

Natural language processing (NLP) shows promise as a means to automate the labelling of hospital-scale neuroradiology magnetic resonance imaging (MRI) datasets for computer vision applications. To date, however, there has been no thorough investigation into the validity of this approach, including determining the accuracy of report labels compared to image labels as well as examining the performance of non-specialist labellers. In this work, we draw on the experience of a team of neuroradiologists who labelled over 5000 MRI neuroradiology reports as part of a project to build a dedicated deep learning-based neuroradiology report classifier. We show that, in our experience, assigning binary labels (i.e. normal vs abnormal) to images from reports alone is highly accurate. In contrast to the binary labels, however, the accuracy of more granular labelling is dependent on the category, and we highlight reasons for this discrepancy. We also show that downstream model performance is reduced when labelling of training reports is performed by a non-specialist. To allow other researchers to accelerate their research, we make our refined abnormality definitions and labelling rules available, as well as our easy-to-use radiology report labelling tool which helps streamline this process.

\end{abstract}
\section{Introduction}
Deep learning-based computer vision systems hold promise for a variety of applications in neuroradiology. However, a rate-limiting step to clinical adoption is the labelling of large datasets for model training, a laborious task requiring considerable domain knowledge and experience. Following recent breakthroughs in natural language processing (NLP), it is becoming feasible to automate this task by training text classification models to derive labels from radiology reports and to assign these labels to the corresponding images \cite{shin2017classification}\cite{wood2020automated}\cite{zech}\cite{GARG20192045}. To date, however, there has been no investigation into the general validity of this approach, including determining the accuracy of report labels compared to image labels as well as assessing the performance of non-specialist labellers.\\

In this work we draw on the experience of a team of neuroradiologists who labelled over 5000 magnetic resonance imaging (MRI) neuroradiology reports as part of a project to build a dedicated deep learning-based neuroradiology report classifier. In particular, we examine several aspects of this process which have hitherto been neglected, namely (i) the degree to which radiology reports faithfully reflect image findings (ii) whether the labelling of reports for model training can be reliably outsourced to clinicians who are not specialists (here we examined whether the performance of a neurologist or radiology trainee (UK registrar grade; US resident equivalent) is similar to that of a neuroradiologist) (iii) the difficulty of creating an exhaustive and consistent set of labelling rules, and (iv) the extent to which abnormalities labelled on the basis of examination-level reports are detectable on MRI sequences likely to be available to a computer vision model.\\

Overall, our findings support the validity of deriving image labels from neuroradiology reports, but with several important caveats. We find that, contrary to basic assumptions often made for this methodology, radiological reports are often less accurate than image findings. Indeed, certain categories of neuroradiological abnormality are inaccurately reported. We conclude that, in our experience assigning binary labels (i.e. normal vs abnormal) to images from reports alone is very accurate. The accuracy of more granular labelling, however, is dependent on the category, and we highlight reasons for this discrepancy. \\

We also find that several aspects of model training are more challenging than is suggested by a review of the literature. For example, designing a complete set of clinically relevant abnormalities for report labelling, and the rules by which these were applied, took our team of four neuroradiologists more than six months to complete with multiple iterations, and involved the preliminary inspection of over 1,000 radiology reports. To allow other researchers to bypass this step and accelerate their research, we make our refined abnormality definitions and labelling rules available. We also make our radiology report labelling tool available which helps streamline this manual annotation process. Importantly, we found that even when enabled with the labelling tool and set of abnormalities and rules, report annotation for model training must be performed by experienced neuroradiologists, because a considerable reduction in model performance was seen when labelling was performed by a neurologist or a radiology trainee. \\

\section{Related work}
NLP models have previously been employed to assign image labels in the context of training computer vision models for neuroradiology applications using radiology reports from both computed tomography (CT) \cite{shin2017classification}\cite{zech}\cite{ct_report} and MRI \cite{wood2020automated} examinations. In all cases, classification performance was reported for the primary objective of labelling reports. However, there was no comparison of either the predicted or annotated labels with the images. The closest published work to our paper is therefore a conference abstract highlighting discrepancies between the findings detailed in chest radiograph reports and the corresponding images when labelling a limited set of abnormalities \cite{olatunji2019caveats}.  To the best of our knowledge no such investigation has been performed in the context of neuroradiology, nor have the challenges of creating an NLP labelling tool for neuroradiology applications been described. \\

Previous work has investigated the accuracy of using crowdsourcing to label images in the context of general \cite{turk} as well as medical \cite{cocos-etal-2015-effectively} computer vision tasks. However, we know of no work in the context of neuroradiology which investigates the level of expertise required for accurate manual annotation of reports. Although it might seem obvious that experienced neuroradiologists are required for this task, previous works have instead employed post-graduate radiology and neurosurgery residents \cite{zech} or attending physicians \cite{shin2017classification}\cite{ct_report}, without providing any insight into the possible reduction in labelling accuracy that such delegation may invite. \\

Automated brain abnormality detection using either $\textrm{T}_{2}$-weighted or diffusion-weighted images (DWI) and employing supervised \cite{rezaei2017brain}\cite{radiol} and unsupervised \cite{chen2018unsupervised} deep learning models has previously been reported. However, in each case only a limited set of abnormalities were available during training and testing, and there was no investigation into the range of abnormalities likely to be detected by the computer vision system using only these sequences. In fact, to the best of our knowledge no investigation has determined what fraction of abnormalities are visible to expert neuroradiologists inspecting only a limited number of sequences. Resolving this point could help narrow the architecture search space for future deep learning-based abnormality detection systems.

\section{Data and methods}
The UK’s National Health Research Authority and Research Ethics Committee approved this study. 126,556 radiology reports produced by expert neuroradiologists (UK consultant grade; US attending equivalent), consisting of all adult ($>$ 18 years old) MRI head examinations performed at Kings College Hospital NHS Foundation Trust, London, UK (KCH) between 2008 and 2019, were included in this study. The reports were extracted from the Computerised Radiology Information System (CRIS) (Healthcare Software Systems, Mansfield, UK) and all data was de-identified.  Over the course of more than twelve months, 5000 reports were annotated by a team of neuroradiologists to generate reference standard report labels to train the neuroradiology report classifier described in \cite{wood2020automated} (ALARM classifier). Briefly, each unstructured report was typically composed of 5-10 sentences of image interpretation, and sometimes included information from the scan protocol, comments regarding the patient’s clinical history, and recommended actions for the referring doctor. In the current paper, we refer to these reference standard labels generated on the basis of manual inspection of radiology reports as ‘silver reference standard labels’. Prior to manual labelling, a complete set of clinically relevant categories of neuroradiological abnormality, as well as the rules by which reports were labelled, were generated following six months of iterative experiments involving the inspection of over 1000 radiology reports. The complete set of abnormalities, grouped by category, are presented in the supplemental material.  \\

Three thousand reports were independently labelled by two neuroradiologists for the presence or absence of any of these abnormalities. We refer to this as the `coarse dataset’ (i.e. normal vs. abnormal). Agreement between these two labellers was 94.9\%, with a consensus classification decision made with a third neuroradiologist where there was disagreement. 
Separately, 2000 reports were labelled by a team of three neuroradiologists for the presence or absence of each of 12 more specialised categories of abnormality (mass e.g. tumour; acute stroke; white matter inflammation; vascular abnormality e.g. aneurysm; damage e.g. previous brain injury; Fazekas small vessel disease score \cite{fazekas}; supratentorial atrophy; infratentorial atrophy; foreign body; haemorrhage; hydrocephalus; extra-cranial abnormality). We refer to this as the ‘granular dataset’. There was unanimous agreement between these three labellers across each category for 95.3\% of reports, with a consensus classification decision made with all three neuroradiologists where there was disagreement. \\

We manually inspected 500 images (comprising, on average, 6 MRI sequences) to generate reference standard image labels. We refer to labels generated in this way as ‘gold reference standard labels’. 250 images were labelled for the presence or absence of any abnormality, systematically  following the same criteria as that used to generate the coarse report dataset. Similarly, 250 images were examined and given 12 binary labels corresponding to the presence or absence of each of the more granular abnormality categories.\\

\begin{figure}
\centering
\includegraphics[width=0.83\textwidth]{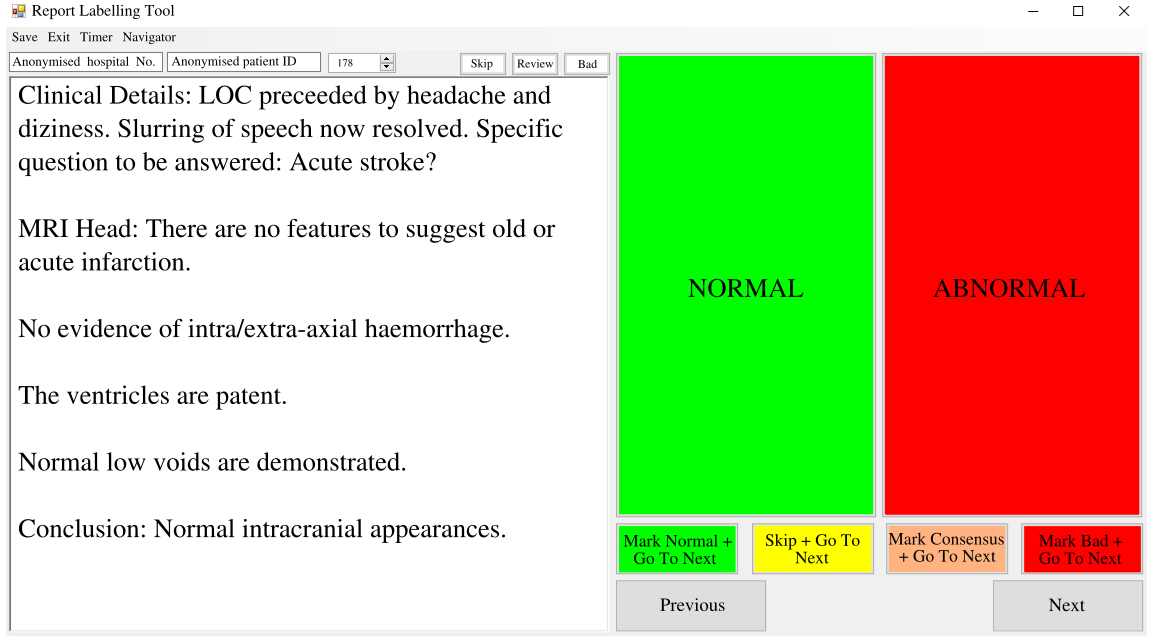}
\caption{Binary report labelling tool for the MR Imaging abnormality Deep learning Identification (MIDI) study. The example report should be marked as normal.} \label{bin_app}
\vspace{-1em}
\end{figure}

\begin{figure}
\vspace{-1em}
\centering
\includegraphics[width=0.83\textwidth]{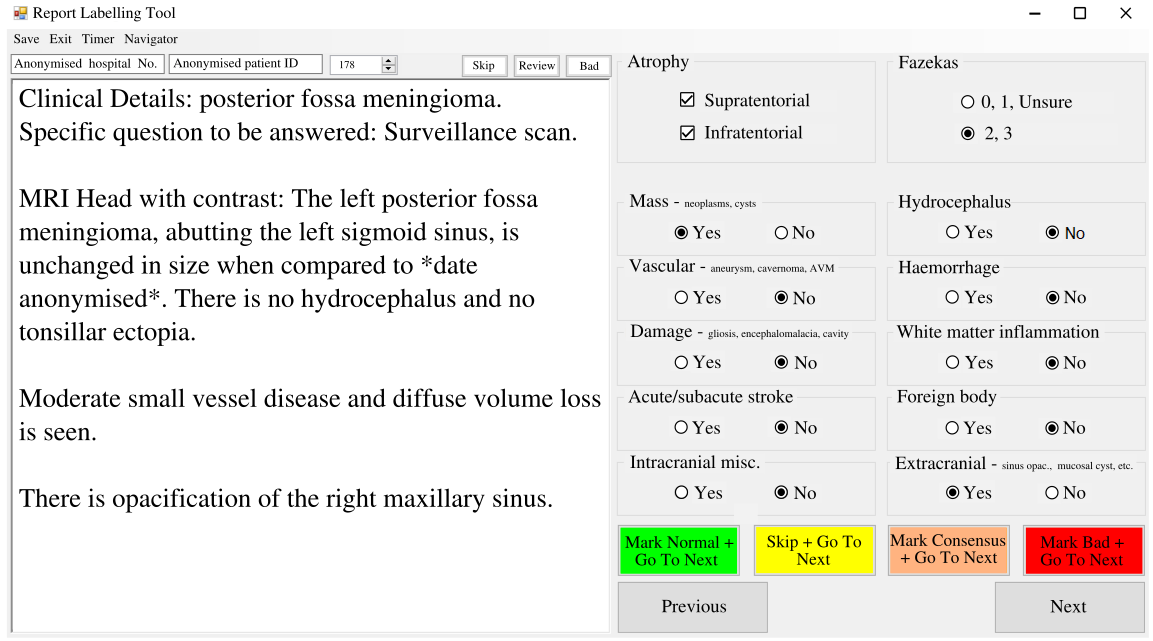}
\caption{Granular report labelling tool for the MIDI study. The correct labels for this example report have been selected.} \label{gran_app}
\end{figure}

Our team designed a complete set of clinically relevant categories capable of accurately capturing the full range of pathologies which present on brain MRI scans. The aim here was to try and emulate the behaviour of a radiologist in the real world, guided by the need for clinical intervention for an abnormal finding. To help other researchers bypass this step, and to encourage standardization across research groups of abnormality definitions, we make our abnormality categories, as well as all clinical rules, available in the supplemental material. Our manual labelling campaign was considerably aided by our development of a dedicated labelling app. This tool allows easy visualisation and labelling of reports through a graphical user interface (GUI), and includes functionality for flagging difficult cases for group consensus/review. Two apps were developed - one for binary labelling (Figure \ref{bin_app}), and one for more granular labelling (Figure \ref{gran_app}) - and we make both available to other researchers at \url{https://github.com/MIDIconsortium/RadReports}.

\section{Results}
\subsection{Impact of annotator expertise}
To assess the level of expertise required to perform manual annotation of reports for training a text classification model, two experiments were performed. \\

First, we compared the coarse labels (i.e. normal vs. abnormal) generated by a hospital doctor with ten years experience as a stroke physician and neurologist, who was trained by our team of neuroradiologists over a six month period, with neuroradiologist-generated labels. The rationale for determining the performance was twofold. Neurologists and stroke physicians frequently interpret reports held on the Electronic Patient Record during patient consultations, therefore it is expected that they would be able to differentiate, and therefore label, normal or abnormal reports accurately. Moreover, given that there are less neuroradiologists than neurologists or stroke physicians, with a ratio of 1:4 in the UK, it is likely to be easier to recruit such physicians to perform such labelling tasks.\\

\begin{figure}
\centering
\includegraphics[width=0.7\textwidth]{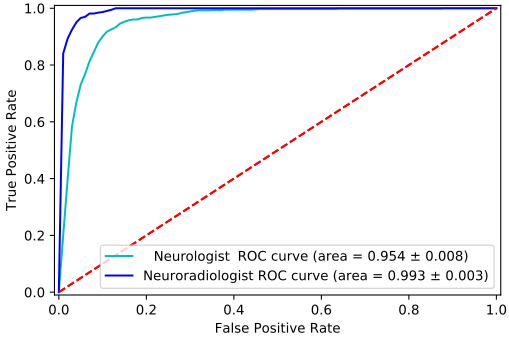}
\caption{ROC curve for a neuroradiology report classifier trained on labels generated by a neurologist (cyan) and a neuroradiologist (blue). The area under the curve (AUC) is shown.} \label{fig1}
\end{figure}

We found a reduction in performance of neurologist labelling when compared to the labels created by an expert neuroradiologist (Table \ref{tab1}). Based on classification and evaluation methodology in \cite{wood2020automated}, the state-of-the-art  ALARM classifier was trained using these neurologist-derived labels and, for comparison, labels generated by a blinded neuroradiologist (Figure \ref{fig1}). The corresponding reduction in classification performance on a hold-out test set of silver reference-standard labels  (i.e. reports with consensus) at an arbitrarily fixed sensitivity of 90\% (Table \ref{classifier_table}) demonstrates the impact of what we have shown to be a sub-optimal labelling strategy.  In summary, there is optimal performance when the classifier is trained with reports labelled by an experienced neuroradiologist.

\begin{table}
\centering
\caption{Labelling performance of a stroke physician and neurologist.}\label{tab1}
\begin{tabular}{|l|l|l|}
\hline
Accuracy (\%) &  Sensitivity (\%) & Specificity (\%)\\
\hline
92.7 &  77.2 & 98.9\\

\hline
\end{tabular}
\end{table}
\begin{table}
\centering
\caption{Accuracy, specificity, and F1 score of a neuroradiology report classifier trained using data labelled by either a neurologist or neuroradiologist operating at a fixed sensitivity of 90\%. Best performance in bold.}\label{classifier_table}
\begin{tabular}{|l|l|l|l|}
\hline
Annotator & Accuracy (\%)  & Specificity (\%) & F1 (\%)     \\
\hline
Neurologist & 89.8 &  89.5 & 75.8\\
\hline
Neuroradiologist & \bfseries 96.4 &  \bfseries 97.7 & \bfseries 90.3 \\

\hline
\end{tabular}
\end{table}

As a second experiment, a 3rd year radiology trainee who was also trained by our team over a six month period to label neuroradiology reports, generated labels for our `granular dataset’. There was a reduction in radiology trainee performance, averaged across all 12 binary labels, when compared to the silver reference standard labels created by our team of expert neuroradiologists (Table \ref{antanas}).  The sensitivity of these labels is clearly too low to be used for model training. 
\begin{table}
\centering
\caption{Labelling performance of a radiology trainee on the `granular dataset', averaged across all 12 binary labels.}\label{antanas}
\begin{tabular}{|l|l|l|}
\hline
 Sensitivity (\%) & Specificity (\%) & F1 (\%)\\
\hline
64.4 &  98.3 & 70.8\\

\hline
\end{tabular}
\end{table}

It is worth highlighting that reliability (inter-rater agreement) and accuracy (performance) should not be conflated for labelling tasks. We demonstrate this in a further experiment where the same neurologist  previously described also generated labels for our `granular dataset'. The Fleiss $\kappa$ score for the radiology trainee and the neurologist averaged over all 12 binary categories was 0.64, which is above the threshold previously employed to establish neuroradiology label reliability\cite{zech}. Substantial inter-rater agreement (commonly taken as $\kappa > 0.6$), therefore, does not necessarily equate to label accuracy as this experiment has shown.

\subsection{Report validation}
To determine the validity of assigning image labels on the basis of radiology reports, the granular labels derived from reports (silver reference standard) were compared to those derived by inspecting the corresponding images (gold reference standard) for 500 cases (Table \ref{granular_stats}).  Although the false positive rate of report labelling is very low for the 12 granular categories of interest, it is clear that the sensitivity of radiology report labelling is category dependent and can be low. On further analysis, we found that insensitive labelling for any given category typically reflects the absence of any reference in the report to that particular category rather than a discrepancy in interpretation. The categories with low sensitivity include hydrocephalus, haemorrhage, extra-cranial abnormalities, and infratentorial atrophy. The reasons for this are discussed below. 

\begin{table}
\centering
\caption{Accuracy of silver reference standard report labels for granular categories when compared to the corresponding gold standard image labels. Categories with sensitivity $>$ 80\% in bold.}\label{granular_stats}
\begin{tabular}{|l|l|l|l|l|}
\hline
 Category &  Sensitivity (\%)  &  Specificity (\%) &  F1 (\%)\\
\hline
\bfseries Fazekas & \bfseries90.5 & 95.6 & 93.2\\
\hline
\bfseries Mass & \bfseries97.9 & 93.6 & 95.9\\
\hline
\bfseries Vascular & \bfseries83.3 & 88.4 & 86.5\\
\hline
\bfseries Damage & \bfseries82.4 & 92.7 & 87.8\\
\hline
\bfseries Acute Stroke &  \bfseries94.4 & 99.5 & 94.4\\
\hline
Haemorrhage & 69.2 & 99.6  & 78.3 \\
\hline
Hydrocephalus & 70.0 & 99.6 & 77.8\\
\hline
\bfseries White Matter Inflammation & \bfseries95.6 & 100 & 97.7\\
\hline
\bfseries Foreign Body  & \bfseries100.0 & 99.6 & 96.6\\
\hline

Extracranial abnormality  & 60.0 & 94.7 & 54.5\\
\hline
\bfseries Supratentorial Atrophy  & \bfseries100 &  94.6 & 76.9\\
\hline
Infratentorial Atrophy  & 77.7 & 94.3  & 54.5\\
\hline
 Macro-average  & 85.1 & 96.0 & 82.8\\
\hline
\end{tabular}
\end{table}

Importantly, silver standard binary labels indicating the presence or absence of any abnormality in a report (i.e. normal vs. abnormal) were accurate when compared to the image (gold reference standard label) (Table \ref{binary_perf}).  

\begin{table}
\centering
\caption{Accuracy of silver reference standard report labels for binary categories (i.e. normal vs  abnormal) relative to the corresponding gold standard image labels.}\label{binary_perf}
\begin{tabular}{|l|l|l|l|}
\hline
Category & Sensitivity (\%)  & Specificity (\%) & F1 (\%)\\
\hline
Normal vs. abnormal & 98.7 & 96.6 & 98.5\\

\hline
\end{tabular}
\end{table}
\subsection{MRI sequences and abnormality visibility}
In another experiment we examined the utility of assigning examination-level labels derived from radiology reports to  different MRI sequences. In general, neuroradiology reports detail findings from multi-modality (i.e. multiple MRI sequences) imaging examinations, with individual sequences providing complementary information to discriminate specific tissues, anatomies and pathologies. For example, the signal characteristics of blood changes over time, the rate of which is sequence dependent. Therefore analysis of images from multiple sequences allows the chronicity of a haemorrhage to be deduced. Assigning the same label to all images in a multi-modality examination can confound computer vision classification if a model isn't optimised to take as its input the individual sequence from which a particular examination-level label was derived. Therefore, we wished to determine whether a minimal number of sequences would be sufficient for use with report-derived labels.  At our institution, axial $\textrm{T}_2$-weighted and DWI images are typically obtained for routine image review, with over 78\% of patients receiving both images during an examination. We sought to determine what fraction of abnormalities are visible to a neuroradiologist inspecting only the $\textrm{T}_2$-weighted and DWI images. Binary labels (i.e. normal vs. abnormal) for 250 examinations were generated by inspecting only these sequences, and compared to labels derived from all available sequences for the same examinations. The agreement between these two labels was 97.8\%, showing that these two sequences would be sufficient for use with report-derived labels for most abnormality detection tasks. Examples of the wide range of abnormalities identified on the basis of $\textrm{T}_2$-weighted and DWI imaging appear in the supplemental material, along with reports describing abnormalities which weren't visible on either of these two sequences.

\section{Discussion}
In this work we have examined several assumptions which are fundamental to the process of deriving image labels from radiology reports. Overall, our findings support the validity of deriving image labels from neuroradiology reports. In particular, assigning binary labels (i.e. normal vs abnormal) to images from reports alone is highly accurate and therefore acceptable. Until now this has been assumed but has not been thoroughly investigated. The accuracy of more granular labelling, however, is dependent on the category. For example, labelling of acute stroke, mass, neuro-degeneration, and vascular disorders, is  shown to be accurate.\\

The low labelling accuracy seen in some granular labelling categories is a result of low sensitivity. Low sensitivity typically reflects the absence of any reference in the report to that particular category rather than a discrepancy in interpretation.  A qualitative  analysis by our team of neuroradiologists has determined several reasons for low sensitivity in some categories. \\

First, in the presence of more clinically important findings, neuroradiologists often omit descriptions of less critical abnormalities which may not necessarily change the overall conclusion or instigate a change in the patient’s management. For example, on follow-up imaging of previously resected tumours, we have found that the pertinent finding as to whether there is any progressive or recurrent tumour is invariably commented on. In contrast, the presence of white matter changes secondary to previous radiotherapy appears less important within this clinical context. If unchanged from the previous imaging, a statement to the effect of “otherwise stable intracranial appearances” is typical in these cases. \\

A second source of low sensitivity is the observation that radiology reports are often tailored to specific clinical contexts and the referrer. A report aimed at a neurologist referrer who is specifically enquiring about a neurodegenerative process in a patient with new onset dementia, for example, may make comments about subtle parenchymal atrophy. In contrast,  parenchymal volumes may not be scrutinised as closely in the context of someone who has presented with a vascular abnormality, such as an aneurysm, and a report is aimed at a vascular neurosurgeon. 
Both sources of low sensitivity mentioned above often reflect a “satisfaction of search error” where the radiologist has failed to appreciate the full gamut of abnormalities. After identifying one or two abnormalities the task may appear complete and there is less desire to continue to interrogate the image \cite{berbaum_franken_caldwell_schartz_madsen_2018}. It is also noteworthy that abnormalities which are identified by the neuroradiologist by chance may be judiciously omitted from the report on a case by case basis when such “incidentalomas” are thought to be of little consequence. Because of these sources of low sensitivity,  labelling categories of abnormality from radiology reports remains challenging for haemorrhage (note that acute haemorrhage is typically detected by CT; MRI reports were often insensitive to those haemorrhages associated with non-critical findings such as micro-haemorrhages), hydrocephalus, extracranial abnormalities and infratentorial atrophy.\\

In addition to examining the accuracy of  radiology reports compared to image findings, we have also demonstrated that most abnormalities typical of a real-world triage environment are picked up using only  $\textrm{T}_{2}$-weighted and DWI sequences. This observation may help narrow the architecture search-space for future deep learning-based brain abnormality detection systems, and allow a more accurate comparison of model performance across research groups. However, there are certain abnormalities which may not be visible on these sequences. For example, the presence of microhaemorrhages or blood breakdown products (hemosiderin), are sometimes only visible on gradient echo ($\textrm{T}_{2}^*$-weighted) or susceptibility weighted imaging (SWI) \cite{chiewvit}. Furthermore, foci of pathological enhancement on post contrast $\textrm{T}_{1}$-weighted imaging can indicate underlying disease which may not be apparent on other sequences. Therefore, whilst we have shown that using $\textrm{T}_{2}$-weighted and DWI sequences alone allows almost all abnormalities to be identified visually, and that plausibly this will translate to efficient computer vison training tasks, it is important to be aware that there are potential limitations. \\

We briefly discuss several logistical aspects of the report labelling process which were not covered by our more quantitative investigations. Our team designed a complete set of clinically relevant categories capable of accurately capturing the full range of pathologies which present on brain MRI scans. The aim here was to try and emulate the behaviour of a radiologist in the real world, guided by the need for clinical intervention for an abnormal finding. This process, however, was more onerous than is often presented in the literature, requiring the inspection of over 1000 radiology reports by our team of experienced neuroradiologists over the course of more than six months before an exhaustive and consistent set of abnormality categories, as well as the rules by which reports were to be labelled, could be finalised. The rules and definitions constantly evolved during the course of the practice labelling experiments. To allow other researchers to bypass this step and accelerate their research, we make our refined abnormality definitions and labelling rules available as well as our dedicated labelling easy-to-use app.\\
\section{Conclusion}
We conclude that in our experience, assigning binary labels (i.e. normal vs abnormal) to images from reports alone is highly accurate. Importantly, we found that even when enabled with the labelling tool and set of abnormalities and rules, annotation of reports for model training must be performed by experienced neuroradiologists, because a considerable reduction in model performance was seen when labelling was performed by a neurologist or a radiology trainee.  In contrast to the binary labels, the accuracy of more granular labelling is dependent on the category.

%
%
\bibliographystyle{splncs04}
\bibliography{paper09.bib}

\end{document}


\title{- SUPPLEMENTAL MATERIAL -\\                             
Labelling imaging datasets on the basis of neuroradiology reports: a validation study}
\titlerunning{NLP for neuroradiology image labelling}

\author{David A. Wood\inst{1} \and
Sina Kafiabadi\inst{2} \and
Aisha Al Busaidi\inst{2} \and
Emily Guilhem\inst{2} \and
Jeremy Lynch\inst{2} \and
Matthew Townend\inst{3} \and
Antanas Montvila\inst{2} \and
Juveria Siddiqui\inst{2} \and
Naveen Gadapa\inst{2} \and
Matthew Benger\inst{2} \and
Gareth Barker\inst{4} \and
Sebastian Ourselin\inst{1} \and
James H. Cole\inst{4,5} \and
Thomas C. Booth\inst{1,2}\\
\vspace{1em}
MR Imaging abnormality Deep learning Identification study (MIDI) Consortium}

%
\authorrunning{D. Wood et al.}
%
\institute{School of Biomedical Engineering, King’s College London \and
King’s College Hospital, London, UK \and
Wrightington, Wigan \& Leigh NHSFT \and
Institute of Psychiatry, Psychology \& Neuroscience, King’s College London \and
Centre for Medical Image Computing, Dementia Research, University College London}
\maketitle              

\section{MR Imaging abnormality Deep learning Identification study (MIDI) neuroradiology abnormality classification definitions }

\noindent Preliminary notes:
\begin{enumerate}

    \item Please use both the CLINICAL INFORMATION (immediately above the report) and the REPORT when labelling as both can provide clues as to how to label
    \item The SKIP button is to be used if entirely unsure
    \item The CONSENSUS button is to be used if nuanced and needs decision by group
    \item Anything slightly ambiguous to the labeller should be “skipped” or go to “consensus required” - never guess
    \item  When there is a differential diagnosis, if it is clear the lesion is very much non-specific and is described as A or B it is OK to label both “A” and “B”. However, if the report is clearly leading the reader towards “A” and only mentioning the other differential i.e. “B” as a mere possibility then we label as “A”. An example might be of the report saying something like “enhancement in the resection cavity likely represents normal post-operative appearances, however residual tumour cannot be entirely excluded” – this should be labelled as “damage” and not “tumour” (see category rules below)
    \item Midline shift would typically be seen in the context of another abnormality. The primary pathology should be labelled only e.g. “tumour” or “acute/subacute stroke” 
    \item Similarly, vasogenic oedema would typically be seen in the context of another abnormality. The primary pathology should be labelled only e.g. “tumour”
    \item Post surgical pituitary scans will be put into CONSENSUS box due to the difficulty of determining from report whether there is a cavity (damage) or not unless this is clearly described within the report
    \item Craniotomies \& craniectomies \& burr holes
    \begin{itemize}
        \item Consensus is that as we want a classifier to ignore these so we would NOT label as “extra-cranial abnormality” and instead we ignore it
        \item However, if lots of metalwork was involved e.g. in a cranioplasty (or the occasional craniotomy which still contains some metalwork) artefact, given the extreme MRI signal distortion we felt that this should be labelled as “foreign body”
    \end{itemize}
    \item Cervical spine and other non-head MRIs contained in addition to MRI heads should be ignored

   \item if artefact then use Bad Scan button

\end{enumerate}

\subsection{Fazekas}

\cite{fazekas} gives a classification system for white matter lesions (WMLs):
\begin{enumerate}
    \item Mild - punctate WMLS: Fazekas I

    \item Moderate - confluent WMLs: Fazekas II

    \item Severe - extensive confluent WMLs: Fazekas III
\end{enumerate}

To create a binary categorical variable from this system, if the report was unsure/normal or mild this would be categorized as `0' as this never requires treatment for cardiovascular risk factors. However, if there is a description of moderate or severe WMLs, the report would be categorized as “1” as these cases sometimes require treatment for cardiovascular risk factors.\\

Included as normal are scattered non-specific foci of signal abnormality (unless a more defuse or specific pathology is implied) and minor/minimal/mild small vessel disease. Conversely, those cases which are mild to moderate small vessel disease, confluent, or beginning to confluence are treated as abnormal.

\begin{itemize}
    \item If described as `mild to moderate', then label as MODERATE
    \item ‘Modest’ is labelled as mild
    \item Non-specific white matter dots / foci of signal abnormality, unless a more defuse or specific pathology is implied, is mild
    \item Confluencing small vessel disease labelled as moderate to severe
    \item CADASIL labelled as moderate/severe
\end{itemize}

\subsection{Mass}

All the following are categorized as `1' for mass:
\begin{itemize}
    \item[$-$] Neoplasms
    \begin{itemize}
             \item  infiltrative tumours
   
        \item  extra-axial masses e.g. vestibular schwannoma
        
        \item  tumour debulking or partial resection as this includes cavity plus tumor (labelled as both “damage” and “mass”)
        
         \item pituitary adenomas
        
         \item ependymal / subependymal / local meningeal enhancement in the context of a history of an aggressive infiltrative tumor
        
    \end{itemize}
    
   \item[$-$]  Abscess

\item[$-$] Cysts

\begin{itemize}
    \item  retrocebellar cyst is included but mega cisterna magna is ignored

 \item pineal cysts and choroid fissure cysts

\item Including Rathke’s cleft cysts
\end{itemize}

\item[$-$] Focal cortical dysplasia, nodular grey matter heterotopia, subependymal nodules and subcortical tubers
\item[$-$]  Lipoma

\item[$-$] Chronic subdural haematoma / hygroma (i.e. CSF equivalent)

\item[$-$] Ignore perivascular spaces unless giant
\item Brief surgical planning reports where e.g. GBM was in the indication are labelled as mass
\end{itemize}
We have attempted to emulate the decision making of a neuroradiologist for all categories. Note that a finding that might generate a referral to a multidisciplinary meeting for clarification would be included within the granular category e.g. an arachnoid cyst may be ignored in clinical practice, but we included it in the `mass' granular category as these are sometimes referred by non-experts to a multidisciplinary meeting for expert review. Thus our classification is sensitive to ensure patient safety.\\

Examples of mass-like findings considered normal include Thornwaldt’s cysts and perivascular spaces which are mentioned but where no size indication is given.
\subsection{Vascular}
All the following are categorized as `1'  for vascular:

\begin{itemize}
\item[$-$] Aneurysms
\begin{itemize}
    \item including coiled aneurysms regardless of whether there is a residual neck or not 
\end{itemize}

\item[$-$] Arteriovenous malformation

\item[$-$] Arteriovenous dural fistula

\item[$-$] Cavernoma

\item[$-$] Capillary telangiectasia

\item[$-$] Old / non-specific microhaemorrhages

\item[$-$] Petechial haemorrhage

\item[$-$] Developmental venous anomaly

\item[$-$] Venous sinus thrombosis
\item Vasculitis if associated with vessel changes such as luminal stenosis or vessel wall enhancement
\begin{itemize}
    \item in cases of sluggish flow – if strong suspicion of thrombus include otherwise ignore
\end{itemize}

\item[$-$] Arterial occlusion / flow void abnormality or absence

\item[$-$] Venous sinus tumor invasion (labelled as both “vascular” and “mass”)

\item[$-$] Arterial stenosis / attenuation – Include if abnormal. If constitutional / normal variant ignore.
\item  Ignore 3rd ventriculostomy (unless there is a clear description of related parenchymal injury)
\item In the context of biopsy, mention if there is obvious damage, otherwise ignore

\end{itemize} 

Note that a finding that might generate a referral to a multidisciplinary meeting for clarification has been within this category e.g. developmental venous anomaly may be ignored in clinical practice, but we included it in the “vascular” granular category.\\

Examples of vascular-like findings which are considered normal include descriptions of sluggish flow, flow related signal abnormalities (unless they raise the suspicion of thrombus) and vascular fenetrations.
\subsection{Damage}
All the following are categorized as `1'  for damage:
\begin{itemize}
    \item[$-$] Gliosis

\item[$-$] Encephalomalacia

\item[$-$] Cavity

\item[$-$] If the patient has had a craniotomy or biopsy there is likely damage – however, for example in the case of a burr-hole and drain previously inserted into the extra-axial space, this does not automatically constitute damage

\item[$-$] “Post-operative changes / appearances” include as damage

\item[$-$] Tumor debulking or partial resection as this includes cavity plus tumour (labelled as both “damage” and “mass”)

\item[$-$] Chronic infarct / sequelae of infarct

\item[$-$] Chronic haemorrhage / sequelae of haemorrhage (with / without hemosiderin staining)

\item[$-$] Cortical laminar necrosis
\end{itemize}

Examples of damage-like findings which are considered 'normal' include craniotomy,  burrholes, posterior fossa decompression without complication, and 3rd ventriculostomy

\subsection{Acute stroke}
All the following are categorized as `1' for acute stroke:
\begin{itemize}
    \item[$-$] Acute / subacute infarct (if demonstrating restricted diffusion)
\begin{itemize}
    \item Include if there are other descriptors indicating a subacute nature such as swelling or “maturing infarct” even though restricted diffusion has normalised
\end{itemize}

\item[$-$] Parenchymal post-operative restricted diffusion / retraction injury (labelled as both `damage' and `stroke')

\item[$-$] Chronic infarct / sequelae of infarct should be labelled under `damage'
\item If a single event with small diffusion restricting and non-restricting elements then only label as stroke (rather than damage)
\item Mature, established or old infarcts without other descriptors should be labelled under damage 
\item MELAS if associated with restricted diffusion. 
\item Hypoxic ischemic if associated with restricted diffusion. 
\item Vasculitis (according to description) if associated with acute / subacute infarct
\end{itemize}

\subsection{Hydrocephalus}
All the following are categorized as `1' for hydrocephalus:
\begin{itemize}
    \item Acute
    \item Trapped ventricle
    \item Chronic / stable / improving hydrocephalus (it does not matter whether its compensated or not)
    \item Ventricular enlargement in the context of atrophy should be ignored and only marked under atrophy
    \item Include normal pressure hydrocephalus 
\end{itemize}
\subsection{Haemorrhage}
All the following are categorized as `1' for haemorrhage:
\begin{itemize}
    \item Any acute / subacute haemorrhage – parenchymal, subarachnoid, subdural, extradural
    \item Acute microhaemorrhages / petechial haemorrhages should be labelled under haemorrhage. However if its old label as vascular.  Acute haemorrhagic foci in the setting of acute axonal injury label as haemorrhage
    \item When there is old haemorrhagic / blood breakdown products / haemosiderin label as damage
    \item T1 shortening is likely to represent acute haemorrhage  in the immediate post-surgical resection cavity
    \item Vasculitis if associated with haemorrhage
\end{itemize}

\subsection{White matter inflammation}
All the following are categorized as `1' for white matter inflammation:
\begin{itemize}
    \item MS and other demyelinating lesions like ADEM and NMO
    \item MS lesions with cavitation (low T1 signal) - only label as white matter inflammation (no need for additional damage label)
    \item Research scans that use the phrase “No non-MS features” are included
    \item Inflammatory lesions in radiologically isolated syndrome / clinically isolated syndrome
    \item Focal ‘cortical thinning’ i.e. subcortical / cortical lesion label as damage
    \item PML / IRIS
    \item Leukoencephalopathies - congenital or acquired (including toxic)
    \item Encephalitis / encephalopathy if it involves the white matter (HIV/CMV)
    \item PRES
    \item Osmotic demyelination (central pontine myelinolysis/ extrapontine myelinolysis)
    \item Susac’s
    \item Radiation – if describes white matter abnormality
    \item White matter changes in the context of vasculitis - only mention if clearly attributed to vasculitis. If non-specific then do not label unless meets other criteria.
    \item Amyloid-related inflammatory change / inflammatory amyloid
\end{itemize}

\subsection{Foreign body}
All the following are categorized as `1' for foreign body:
\begin{itemize}
    \item Shunts
    \item Clips
    \item Coils
    \item If lots of metalwork was involved e.g. in a cranioplasty (or the occasional craniotomy causing extreme MRI signal distortion)
    \item If craniotomies are not causing significant artefact, then ignore

\end{itemize}

\subsection{Extracranial}
All the following are categorized as `1' for extracranial:
\begin{itemize}
    \item Total mastoid opacification / middle ear effusions
    \item Complete opacification / obstruction of the paranasal sinuses
    \item Ignore mucosal thickening
    \item If there is clearly a well-defined unambiguous polyp then label as abnormal. If it is 'retention cysts' or 'polypoid mucosal thickening' then ignore. If it is something indistinguishable which could be a retention cyst / polyp then ignore. Anything leading to obstruction - always label as abnormal.
    \item Calvarial / extra-calvarial masses
    \item Osteo-dural defects 
    \item Encephaloceles
    \item Pseudomeningoceles
    \item Extracranial vessel abnormality below the petrous segment e.g. cervical ICA dissection
    \item Lipoma, Sebaceous cyst or any other mass if extracranial 
    \item  Orbital abnormalities (including masses)
    \begin{itemize}
        \item Including optic nerve pathology affecting the orbital segment of the nerve i.e. meningioma
        \item If there is an abnormality of the intracranial segment of the optic nerve /chiasm such as atrophy then label as intracranial misc.
    \end{itemize}
    \item Cases with tortuous optic nerves with no other features are ignored
    \item Eye prostheses and proptosis
    \item Ignore pseudophakia
    \item Bone abnormality e.g. low bone signal secondary to haemoglobinopathy
    \item Basilar invagination. 
    \item However hyperostosis is considered normal and not labelled (attempting to mirror how a normal radiologist would approach these)
    \item Thornwald’s cysts are ignored
    
\end{itemize}
\subsection{Intracranial miscellaneous}
All the following are categorized as `1' for Intracranial miscellaneous:
\begin{itemize}
    \item Cerebellar ectopia
    \item Brain herniation (for example into a craniectomy defect)
    \item Clear evidence of idiopathic intracranial hypertension (prominent optic nerve sheaths, intrasellar herniation)
    \begin{itemize}
        \item Non-specific intrasellar arachnoid herniation / empty sella should be otherwise ignored
        \item Non-specific tapering of dural venous sinuses should be ignored
    \end{itemize}
    \item Spontaneous intracranial hypotension (pituitary enlargement, pachymeningeal thickening, etc)
    \begin{itemize}
        \item If subdural collections present, these should be also noted separately
    \end{itemize}
    \item Cerebral oedema or reduced CSF spaces
    \item Absent or hypoplastic structures such as agenesis of the corpus callosum
    \item Meningeal thickening or enhancement – for example in the context of neurosarcoid or vasculitis 
    \item Enhancing or thickened cranial nerves
    \item Infective processes primarily involving the meninges or ependyma (i.e. ventriculitis or meningitis)
    \item Encephalitis if primarily involves the cortex  (HSV/autoimmune encephalitis)
    \item Excessive or unexpected basal ganglia or parenchymal calcification
    \item Optic neuritis involving the intracranial segments of the optic nerves or chiasmitis 
    \item Adhesions / webs
    \item Pneumocephalus
    \item Colpocephaly
    \item Superficial siderosis
    \item Ulegyria
    \item FASIs / UBOs
    \item Basal ganglia / thalamic changes in the context of metabolic abnormalities
    \item Neurovascular conflict - if clearly normal, ignore. If clear cut description of neurovascular conflict such as compression or distortion of cranial nerves then label as intracranial misc. If unsure put under REVIEW
    \item Band heterotopia and polymicrogyria 
    \item Hypophysitis 
    \item Seizure related changes 
    \item ALS (unless there is mention of significant atrophy – it may be worth reviewing the images)

\end{itemize}
\section{Abnormality visibility on $\textrm{T}_2$ or DWI sequences}
\begin{table}
\centering
\caption{Examples of abnormalities detected from axial $\textrm{T}_2$ and DWI sequences only}\label{visible_examples}
\begin{tabular}{|l|}
\hline
acute infarction, neurosarcoidosis, polymicrogyria\\
hypoxic ischemic brain injury, arachnoid cyst, cerebellar haemangioma,\\
Behcet's vasculitis, thalamic haematoma, high grade glioma, \\
occipital lymphoma, sebaceous cyst, periventricular heterotopia, \\
pontine angle epidermoid cyst, white matter injury, mature infarct,\\
parafalcine meningioma, mucosal retention cyst, cavernoma,\\
neurodegeneration, cerebellar tonsillar ectopia, 
demyelination,\\
white matter inflammation, aneurysm, small vessel
ischaemic change, \\
convexity meningioma, Alzheimer's disease, parietal lobe haemorrhage\\

\hline
\end{tabular}
\end{table}
\subsection{Example reports from cases not visible on T2 or DWI sequences}
\begin{itemize}
    \item There is persisting mild circumferential enhancement of the optic nerve sheaths. Slightly equivocal STIR hyperintensity is seen within the canalicular segment of the left optic nerve. The extraocular muscles and orbital fat are normal in appearance, but there is mild bilateral proptosis. The lacrimal glands are of normal size. \textbf{There remains a thin rim of enhancing tissue along the anteromedial aspect of the right middle cranial fossa. No new areas of enhancement or parenchymal signal change are identified}.   Conclusion: Stable appearances of the right middle cranial fossa enhancement. Equivocal signal change in the left optic nerve.\\
    
    \item  There is no acute infarct. \textbf{There is a tiny focus of susceptibility in the left frontal white matter}, no other haemorrhage related susceptibility effects are shown on gradient echo imaging.  There is generalised cerebral volume loss in keeping with the patient's age. No acute intracranial abnormality is shown.  There is no significant stenosis demonstrated on screening TOF MRA carotid imaging. \\
    
    \item MRI Head with contrast: \textbf{There is enhancement within the internal auditory meatus on both sides. This is non-specific and is consistent with either an inflammatory or neoplastic process}. The intracranial appearances are otherwise  normal.\\
    
    \item MRI Head: T2 Axial, Coronal FLAIR, T1 Sagittal and Diffusion imaging of the Brain. There is no cerebral signal abnormality.  There is no structural abnormality in the line of the trigeminal nerves or their major divisions within the skull base or deep face.  There is left sided trigeminal nerve neurovascular contact at 2-3mm from the brain stem, with some moulding of the trigeminal nerves. There is also some minor neurovascular contact within the cisternal portion of the right trigeminal nerves, without evidence of neural distortion.     In Conclusion:  \textbf{There is bilateral neurovascular contact with some distortion of the left trigeminal nerve within the root entry zone}.\\
    
    \item MRI Head with contrast: \textbf{The plaque of dural enhancement and thickening overlying the left frontal convexity and extending into the superior sagittal sinus is unchanged in size and appearance}.   Stable appearances at the presumed left frontal convexity meningioma.  
    
\end{itemize}
\bibliographystyle{splncs04}
\bibliography{paper09_supp.bib}